# Deep learning for detection of bird vocalisations

*Ilyas Potamitis*

**Authors' Affiliations:**
*Ilyas Potamitis*, Technological Educational Institute of Crete, Department of Music Technology and Acoustics, Crete, Greece

**Corresponding Author:** Dr. Ilyas Potamitis, Assoc. Prof.
**e-mail:** potamitis@staff.teicrete.gr
**phone:** 0030 28310 21900
**postal address**: Dept. of Music Technology & Acoustics, Technological Educational Institute of Crete, E. Daskalaki Perivolia, Rethymno 74100, Crete, Greece.
**web**: https://www.teicrete.gr/mta/en/potamitisstaffteicretegr
**google scholar**: http://scholar.google.com/citations?user=gWZ4dTUAAAAJ&hl=en

*Abstract*— This work focuses on reliable detection of bird sound emissions as recorded in the open field. Acoustic detection of avian sounds can be used for the automatized monitoring of multiple bird taxa and querying in long-term recordings for species of interest for researchers, conservation practitioners, and decision makers. Recordings in the wild can be very noisy due to the exposure of the microphones to a large number of audio sources originating from all distances and directions, the number and identity of which cannot be known *a-priori*. The co-existence of the target vocalizations with abiotic interferences in an unconstrained environment is inefficiently treated by current approaches of audio signal enhancement. A technique that would spot only bird vocalization while ignoring other audio sources is of prime importance. These difficulties are tackled in this work, presenting a deep autoencoder that maps the audio spectrogram of bird vocalizations to its corresponding binary mask that encircles the spectral blobs of vocalizations while suppressing other audio sources. The procedure requires minimum human attendance, it is very fast during execution, thus suitable to scan massive volumes of data, in order to analyze them, evaluate insights and hypotheses, identify patterns of bird activity that, hopefully, finally lead to design policies on biodiversity issues.

# I. INTRODUCTION

Birds use acoustic vocalization as a very efficient way to communicate as the sound does not require visual contact between emitter and receiver individuals, can travel over long distances, and can carry the information content under low visibility conditions, such as in dense vegetation and during night time hours [1]. In this paper we will focus only on sounds produced in the vocal organ of birds (i.e. calls and songs).
The operation of autonomous remote audio recording stations and the automatic analysis of their data can assist decision making in a wide spectrum of environmental services, such as: Monitoring of range shifts of animal species due to climate change., biodiversity assessment and inventorying of an area, estimation of species richness and species abundance, assessing the status of threatened species, and alarming of specific atypical sound events related to potentially hazardous events and human activities (e.g. gun shooting) [2-3].
During the last decade the progress of bioacoustic technology is evident especially in the field of hardware development, particularly of programmable and affordable automatic recording units (ARUs). Modern models are powered by solar energy, equipped with large storage capacity, carry weather-proof normal and ultrasound microphones, and some of them are equipped with wireless transmission capabilities [4].
Pattern recognition of bird sounds has a long history and many pattern recognition approaches [5-16] have been applied to the problem of automatic bird detection and identification. This work puts the emphasis on bird detection at large scales involving hundreds of species and massive volumes of unprocessed real-field recordings gathered by ARU's that record unattended for large periods of time. The benefit in gathering such amounts of data diminishes if we do not possess automatic ways to navigate ourselves inside them, to analyze them, to discover and retrieve what we are looking for in order to evaluate insights and hypotheses. Daring a hard statement: A deluge of recording data without a toolbox to analyze them quickly and efficiently, is close to not having any data at all.
This work focuses on a specific question of bird detection in audio: Is there bird activity in a recording clip? If yes, when did it happen? Can you extract segments for further examination? Although the process we will describe is directly expandable to more refining questions, in this work we investigate bird activity in general and we are indifferent to species' composition. That is, we present a generic bird activity detector (presence/absence of bird vocalizations in a clip). The described approach segments the time-frequency spectrum by outputting a zero everywhere but on the locations of bird vocalizations, therefore allows time-stamping, extraction and retrieval of sound snippets. Once trained, it is very fast in execution, requires minimal human attendance during training and none once operational.

The reported literature on the application of Deep learning networks on bird audio recordings is surprisingly sparse [17]. This work introduces a special type of deep learning networks named auto-encoders and the U-net in particular [18]. Our inspiration of using a U-net to detect bird vocalizations was based on observing the images that U-net was initially proposed to segment: neuronal structures in electron microscopy images. To our point of view, the cells under the microscope show visual resemblance to spectral blobs in the audio spectrogram. The U-net architecture consists of a contracting path to capture context around the blobs that ends to a bottleneck and subsequently, a symmetric expanding path that enables the determination of a binary mask imposed on the picture that finally allows localization of cells (or spectral blobs in our case). In [18] a network had been trained end-to-end from very few images and outperformed the prior best method, winning by a large margin including other types of convolutional networks on a cell tracking challenge competition. The use of very small amount of images needed, makes it ideal candidate for small recording databases as in our case here. In the original application of this particular deep learning network, the input in the network during the training process was an image and the output was a binary presence/absence mask of the same dimensions as the input image that localized cells. The training data in [18] are extracted manually. This manual tagging process would be also possible in audio spectrograms (see e.g. [14] as an example of a different to ours approach that requires manual tagging of spectral blobs), but would be highly unpractical for bird vocalizations appearing in an abundance of data. We need a generic tool, as is the case in [19], to segment bird vocalizations *without* requiring the manual, laborious tagging of thousands of audio spectrograms to their detail.

In this work, we use a modified version of [6] to extract automatically the mask of the spectrogram of a bird recording. The training set is composed of spectrogram figures of bird recordings as well as recordings void of any bird activity and their corresponding binary masks. During training, the spectrogram which is a 2D representation is presented as input, and the mask is presented as output, whereas the network learns the mapping in-between them. In the real world picked up by the ARUs, the recordings are almost never truly silent. They include other types of audio events such as rain, wind, mechanical noises, insects, etc. During training, the binary mask of the spectrograms corresponding to recordings that do not include bird vocalizations are mapped to an empty figure. Therefore, the network is forced to learn, by an iterative procedure, the mapping of vocalizations to binary blobs and to suppress everything else.

This paper is organized as follows: In Section II we discuss about bird vocalizations from the signal processing point of view. In Section III we present the description of various versions of U-net as applied to this particular task. In Section IV we present evaluation results on the segmentation efficacy of the U-net on spectrogram images. In Section V we visualize and discuss the predicted outcome of the net on randomly selected audioscenes. Finally we conclude this work in Section VI.

## II. Bird Vocalizations and the Spectrogram

Bird calls usually refer to simple frequency patterns of short monosyllabic sounds. While all birds emit calls, although with different variability and frequency, only some birds also produce songs. In difference to calls, songs are longer, acoustically more complex, and often have a modular structure [1-3].

The spectrogram –also called Short-time Fourier transform- is the outcome of a number of processing stages imposed on audio. The sampled data in the time domain stored in the ARUs, are decomposed into overlapping data chunks that are windowed. Each chunk is subsequently Fourier transformed and the magnitude of the frequency spectrum of each data-chunk is derived. Each spectrum vector corresponds to a vertical line in the image; a measurement of magnitude versus frequency for a specific moment in time. These spectrum vectors are placed side by side to form the spectrum image. An audio scene can be treated as an image through its spectrogram. Acoustic events appear as localised spectral blobs/patches on a two-dimensional matrix (see Fig. 1). The structure of these blobs constitutes the acoustic signature of the sound and is used as a biometric queue to reveal evidence of identity of the source is several audio, pattern recognition applications.

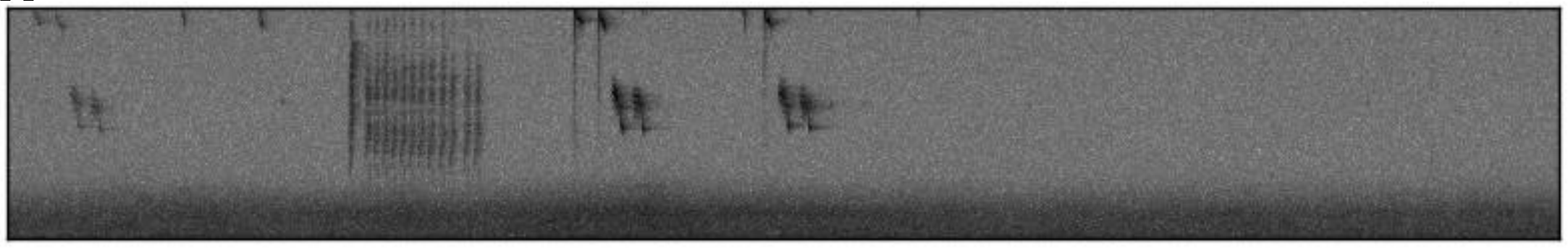

Fig. 1. Spectrogram of bird calls in the presence of strong wind. Audio events stand out as patches of intense colouring.

The audio spectrograms are grey-coloured 1-channel images fed to the input-layer of the U-net. The image fed to the output layer during training is the segmentation of the spectrogram of the input image. For reasons explained in the introduction we need an automatic segmentation method of the spectrogram. In this work we make use of a slightly modified segmentation method of audio spectrograms to derive the masks of audio events, originally described by Lasseck M. in [6]. In this method, the recording is firstly amplitude normalized. Then, morphological operations are applied on the image. These operations derive masks of spectral blobs by connecting regions of high amplitude that correspond to calls or phrases and eliminate small regions of high amplitude that cannot belong to animal vocalizations because they are too short in time and frequency. All these different processing stages are standard in image processing and have as a result the removal of background and extraction of spectral blobs [20]. The binary mask as introduced in [6] is composed of the following stages (see Fig. 2):

1. Removal of Noise: This segmentation method is very effective in handling noise variation due to this image enhancement stage. At this stage, a pixel – corresponding to a time-frequency segment - can potentially become part of a vocalization patch only if it stands out of the median calculated along the frequency axis *and* along the time axis otherwise it is attributed to noise and is set to zero.
2. Binary Closing and Dilation enlarges bright regions and shrinks dark regions.
3. Median filtering: removes ‘salt-and-pepper’ type of residual spectral artifacts.
4. Removal of small objects. Removes small segments that are too small to be bird calls-syllables
5. Finally, one derives a binary mask by labelling as a single segment all connected segments exceeding a certain spatial extension.

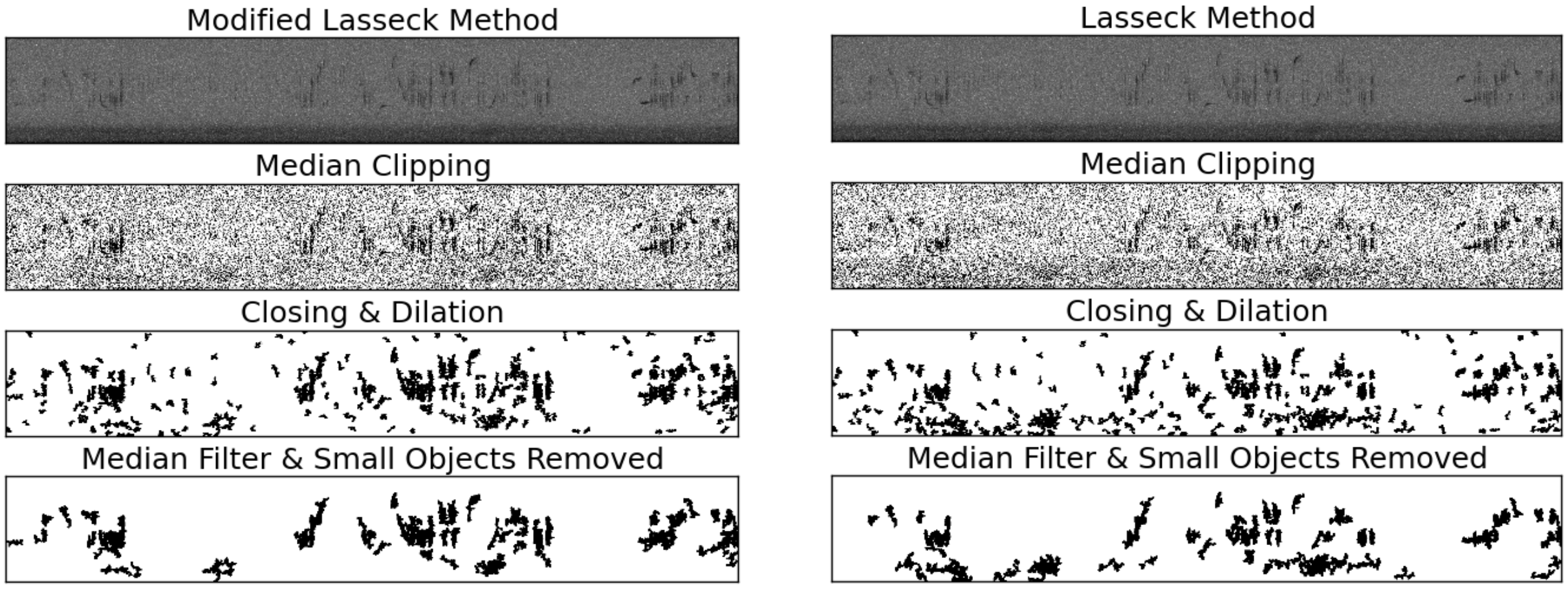

Fig. 2. (Left) segmentation based on a modified method of Lasseck in [6], (Right) Segmentation method based on the original version in [6]. The spectrogram of a bird recording is led through the processing stages of median clipping, morphological operations involving closing & dilation and finally median filtering to a binary mask of presence/absence of audio activity in a spectral patch.

The segmentation method we described is very effective when discarding acoustic events that are distributed over the spectrogram in way that they have lower pixel intensity than the blobs that belong to vocalizations. It is also able to discard acoustic events that fill the whole recording from end to end as in the case of steady state interferences and Cicadas songs. In this work, we modify slightly this method by applying it independently to the high and low frequency parts, thus deriving different segmentation thresholds for the crucial Step-1 of the method. The 256 freq. bins spectrum is split in two parts (the higher 128 spectral bins and the lower frequency part the rest 128 bins) and the rules 1-4 are applied independently to each part with different thresholds. We have found this procedure to produce slightly better enhancement results, and in some cases, where the lower part is contaminated by strong noise, significantly better. The employment of independent thresholds for high/low frequencies result to a mild treatment of upper frequencies that in some cases preserves better feeble vocalizations and spectral detail. Moreover, it effects heavier suppression of low-pass noises that contaminate field recordings in the lower part of the spectrum. After the morphological operations are applied the picture of the spectrogram is made binary in order to mark the masks of the spectral blobs (see Fig. 3-4).

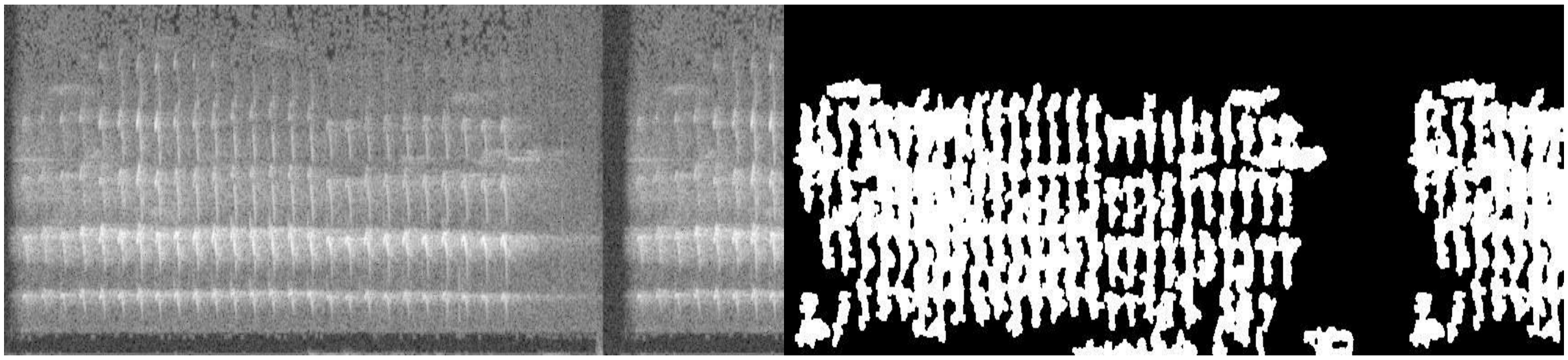

Fig. 3. Bird vocalization. (Left) Spectrogram, (Right) Binary mask produced by a modified version of [6]. Images left and right have the same dimensions and are presented simultaneously to the input and output layer of the U-net. The Deep network learns to map unseen spectrum figures such as the left one to the corresponding binary masks in the right, and, therefore achieve spectral segmentation of audio events.

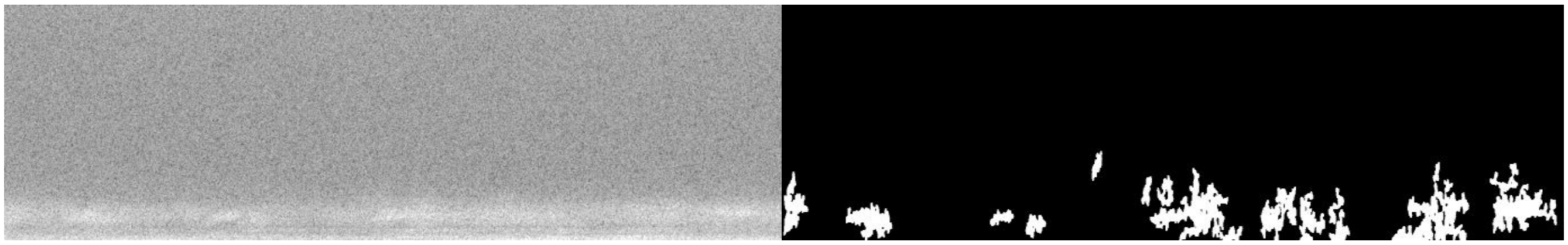

Fig. 4. An audio-scene void of bird activity (wind and background traffic noise). (Left) Spectrogram, (Right) Binary mask corresponding to the spectrogram on the left produced by [6]. Note that the training binary masks of non-bird recordings -as the one in the right figure- are not used during training, instead they are mapped to empty figures (i.e. figures holding only zeros appearing as a uniformly black images).

*Dataset*

Currently, a large number of amateur users and professional providers of data upload and annotate species in various databases such as the Xeno-canto collaborative database (http://www.xeno-canto.org/), which involves more than 140.000 audio records covering 8700 bird species observed all around the world thanks to the active work of more than 1400 contributors. As we need a generic bird detector we downloaded from Xeno-canto 917 bird recordings at random. This amount of data constitutes the training set.

The non-bird corpus, was recorded using automatic recording units (Song Meter SM2, Wildlife Acoustics ®) placed at Lake Vouliagmeni (37˚48’28”N, 23˚47’08”E; c. 10 m a.s.l.), in the Natura 2000 area

"Hymettus – Kaisariani – Lake Vouliagmeni" (GR3000006) at the eastern periphery of the Greek capital Athens. We selected 522 recordings that were visually inspected to be void of bird vocalizations.

The recordings in both corpora were down-sampled at 16000 Hz, converted to monophonic and truncated to the first 15 seconds. Data are not processed or screened in any other way (e.g. band-pass filtering) and, therefore, include all kinds of acoustic degradations typically encountered in nature. Note that the bird recordings in Xeno-Canto often include other audio sources as well (e.g wind or rain audio). The segmentation method will not discern the origin of the sound and will include their blobs in the segmentation to a various extent (see Fig 4 for an example). This is the price to pay for unattended segmentation. However, the network will learn to mitigate this error because the recordings with background noise are mapped to zero (and not to the noisy masks) and therefore, iteratively and in time the network learns to suppress spectral patches not originating from bird activity (see Discussion of results section). The only manual effort is to select by visual inspection of the spectrogram, a folder containing recordings of birds and another folder that containing anything but bird vocalizations. During training the recordings are shuffled but we need to keep them separate to ensure that the masks of spectrograms coming from non-bird activity are mapped to zero.

The FFT has a size of 512 samples and we use 75% overlap and a Hanning window during spectrogram construction. The final dataset consists of images of 256x768 analogous to the number of frequency bins (i.e. half of the 512 samples of the FFT) and to the number of frames. The logarithm of the magnitude of the spectrogram, that is: $20*\log_{10}$(Magnitude Spectrogram) serves as the final image. Logarithmic compression was found beneficial as it compressed the large differences in amplitude encountered in magnitude spectral regions that could lead to poor representation of an image.

## III. Deep Learning as applied to spectrogram images

Being an autoencoder, the U-net does not require thousands of annotated training samples. The basic structure of the deep network receiving 256x768 images in the input and output layers during training and outputting 256x768 images during testing. In Fig. 5, one can see the basic structure of the U-net deep autoencoder and in the Appendix, three variations of it depending on the number of layers one employs and a version with the more advanced and award-winning, inception blocks [21].

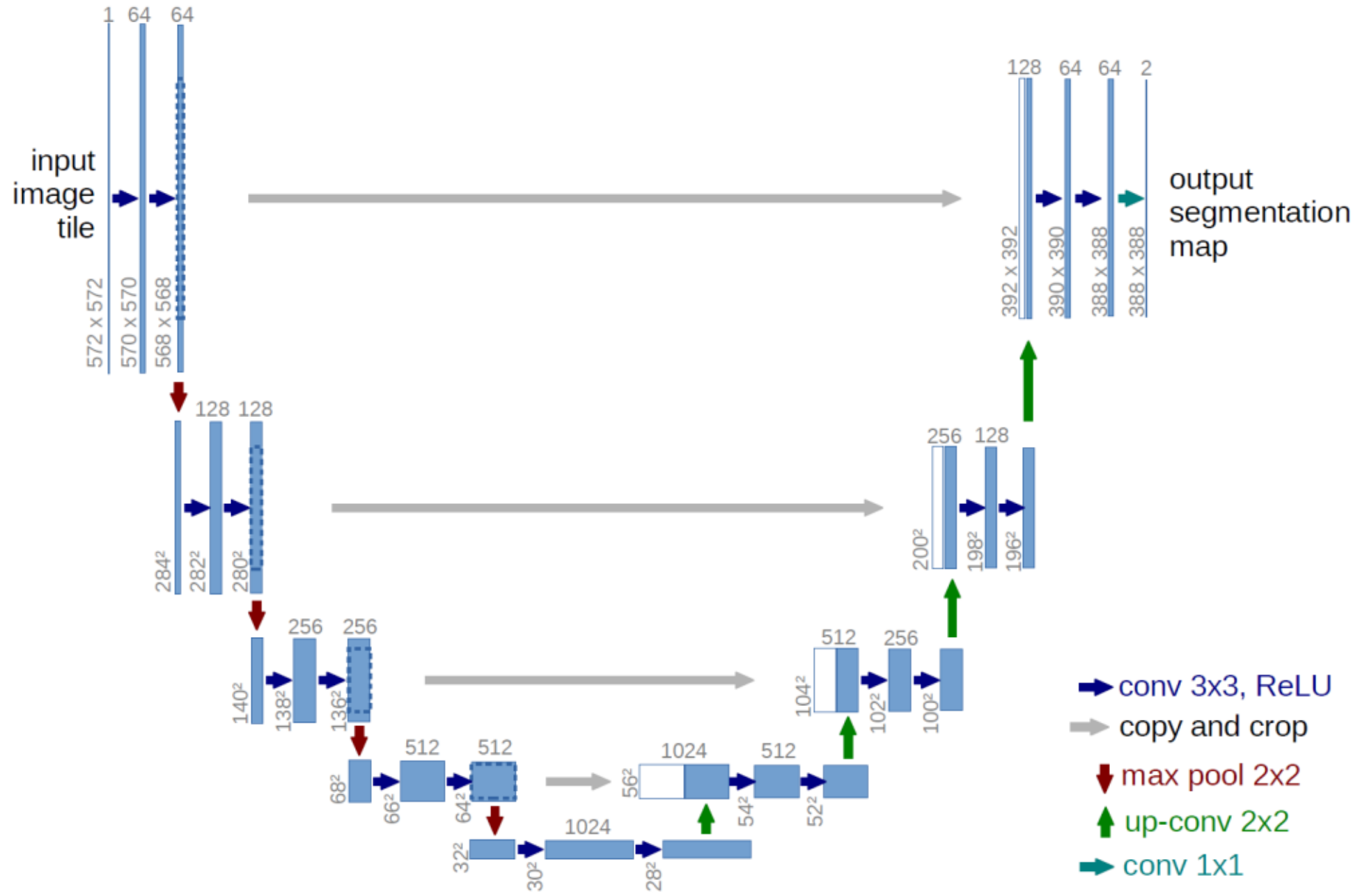

Fig. 5 The structure of the original U-net (after Ronneberger O. et al as presented in [18]). The number of convolutional-pooling layers or the inclusion of processing blocks such as these encountered in Inception [21] and RESNET networks [22] can lead to variations of the same net.

All three methods follow the same data augmentation procedure that contributes significantly to the final accuracy. In the case of the spectrogram one has to be careful of what augmentation methods one applies. The reason for this, is that, deep learning is designed to capture general visual objects that can be anywhere in a photo (e.g. a cat can be located anywhere in a picture and can be flipped in either axis without losing its identity). Moreover a cat in a picture can be large or small depending on the distance from the camera or its focus. This is not the case with spectral segments of bioacoustics data. As specific species cannot vocalize anywhere in the spectrogram. In fact, once the species is known the locations of the vocalizations in the spectrogram are more or less predictable and vice versa. Within-species variation is to be expected but the variation is not to the extent of free moving around the spectrogram, it is rather a deformation in shape. Therefore, horizontal and vertical flipping is not a proper image transformation during data augmentation as applied to bird vocalizations. On the other hand, data augmentation for audio is more than necessary because the specific manifestation of an audio event is just one of the myriad ways it could appear as the horizontal axis is time and a bird can start and stop anywhere in a recording. This fact, together with the chance of co-existing with other audio sources makes data augmentation for spectrograms a necessity.

In this work we did not exhaust the augmentation possibilities and we only used small random rotations between -5 and 5 degrees as well as random displacements at the maximum of the 10% of the image. Although this is a large displacement to apply in order to accommodate within-species vocal variations, we decided to use it and proved beneficial probably because we make a global bird detector and not one dedicated to a specific species.

The simple U-net needs 0.06 sec to process a spectrogram representing 13.79 sec of audio on a single TITAN-X GPU (see Evaluation Section). This entails that it can process 1 hour of recordings in 15.66 seconds or 100 hours in 26.10 minutes. Other accurate classification procedures applied to bird recognition as in [6], and [14-15] do not scale well as the number of recordings increases to the order of tens- to hundreds of thousands 15 sec clips as they need to extract first the features of the test dataset before they classify it. Moreover, classical state-of-the –art classification techniques such as Support Vector Machines and Random Forests do not show, for the time being, to make full use of the parallel processing capabilities of GPUs in the way deep networks do. Note also that this speed is achieved using a single TITAN-X GPU and would improve significantly in a multi-GPU setting.

## IV. Evaluation

We trained the detection framework in terms of the mean Dice coefficient loss function. The Dice coefficient can be used to compare the pixel-wise agreement between a predicted segmentation and its corresponding ground truth. The formula is given by:

$$\frac{2*|X \cap Y|}{|X|+|Y|}$$

Where, X is the predicted set of pixels and Y is the ground truth. The Dice coefficient is the quotient of similarity and ranges between 0 and 1. It can be viewed as a similarity measure over sets. The loss function is just the minus of the Dice coefficient with the additions of a smoothing factor inserted in the denominator. The score in Table I is the mean of the Dice coefficients of images in the evaluation set.

| U-net | Dice Coefficient | Train time (h) | Pred. time/im (sec) |
|---|---|---|---|
| Simple U-net | 0.71 | 5 | 0.06 |
| Enlarged U-net | 0.74 | 7 | 0.16 |
| Inception blocks | 0.65 | 10 | 0.79 |
| | | | |

Table I. All U-net versions are trained with 60 epochs on the same dataset. The Inception block converges slower due to the small batch-size necessary to avoid memory overflow but finally achieves better results after a large number of epochs.

## V. Discussion of Results

In this paragraph we merely comment on prediction results concerning audioscenes we found interesting. All test audiodcenes were excluded from the test set. In Fig. 6 one can see that the lower frequencies of the spectrum (Fig. 6 bottom) are quite heavily suppressed while the structure of the song is retained. High frequency blobs that usually have lower energy are adequately preserved. The predicted blobs are smoother compared to segmentation method of [6] (see Fig. 7).

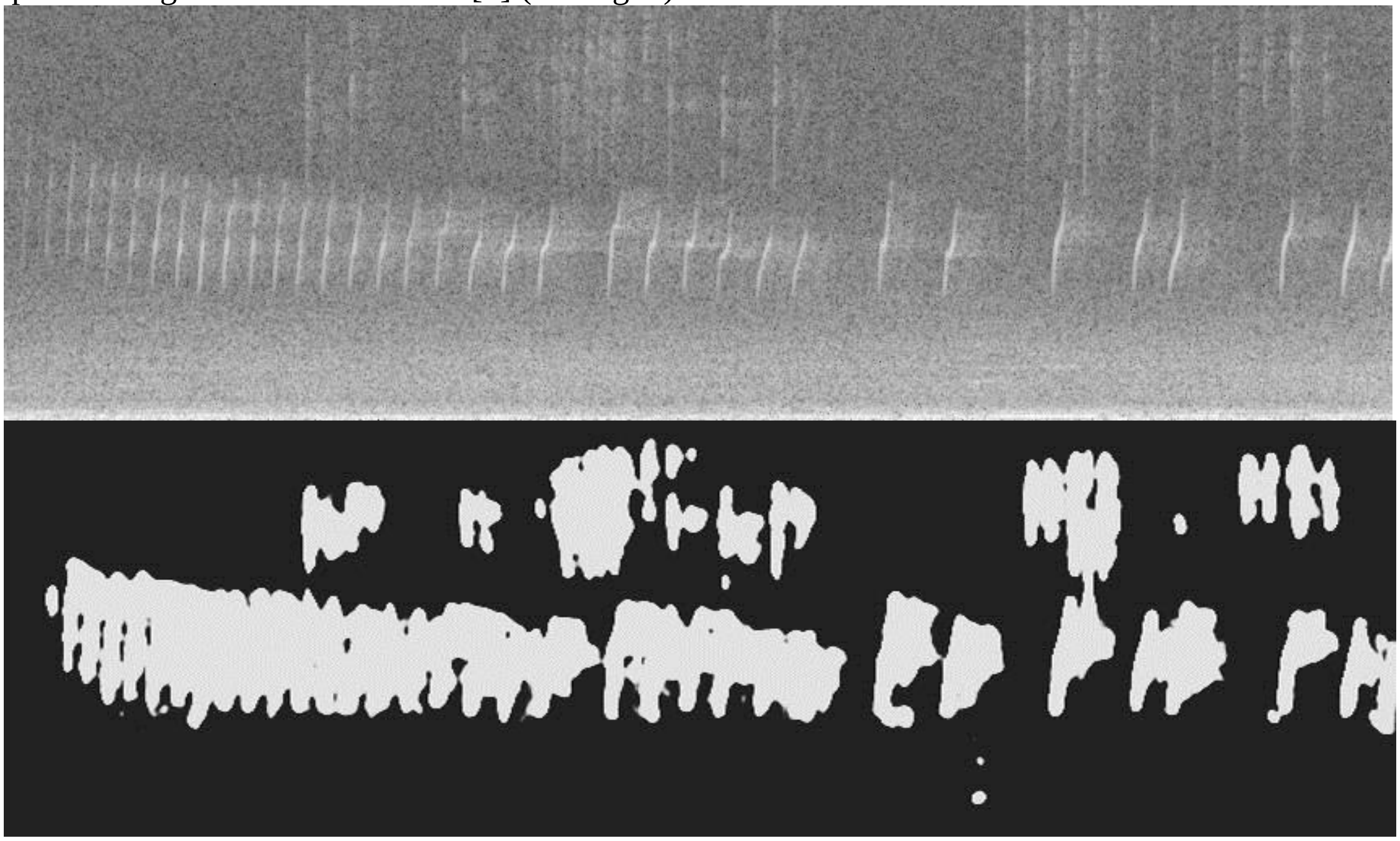

Fig. 6 An audio-scene void of bird activity (wind and background traffic noise). (Left) Spectrogram, (Right) Binary mask. Note that the training binary masks of non-bird recordings as the one in the right are not used during training, instead they are mapped to empty figures (i.e. figures holding only zeros that would appear as a uniformly black image).

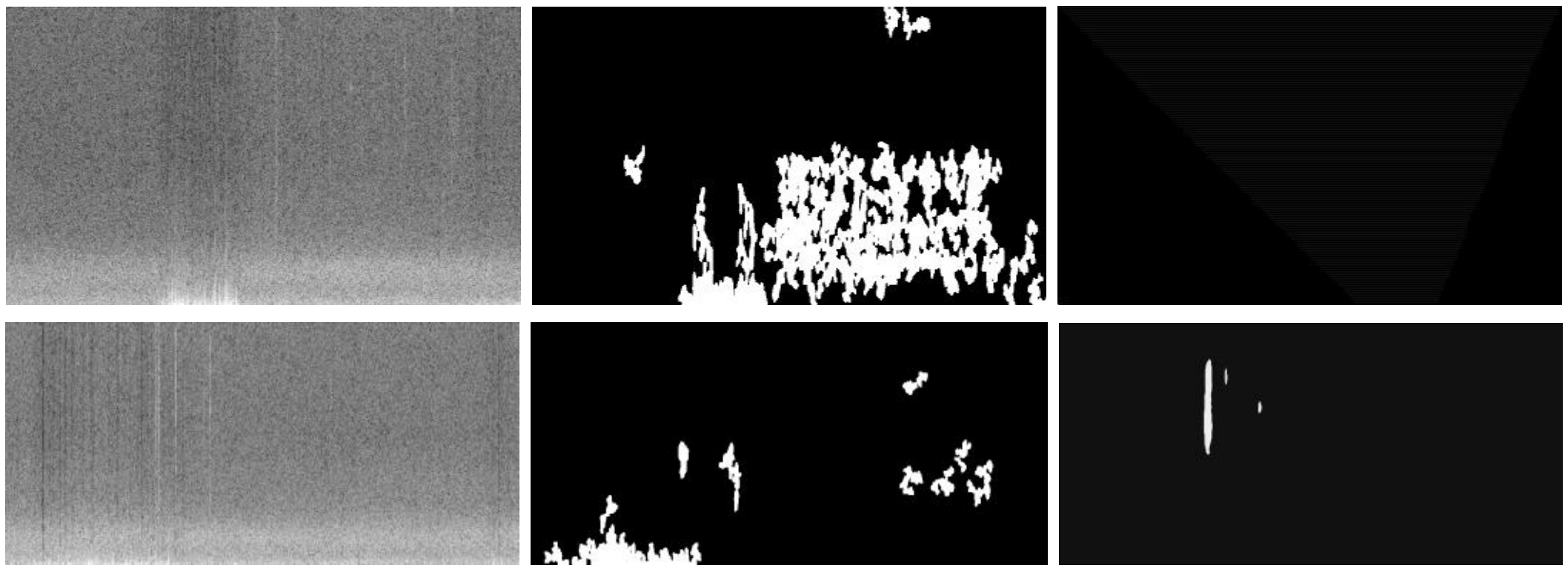

Fig. 7 (Top row). An audio-scene void of bird activity (very strong wind and background traffic noise). (Left) Spectrogram, (Middle), Segmentation using modified [6], (Right) Predicted Binary mask using U-net2. (Bottom row). An audio-scene void of bird activity (very strong wind and mechanical noises). (Left) Spectrogram, (Middle), Segmentation using [6], (Right) Predicted Binary mask using U-net2. Note the effect off directing abiotic sound scenes to empty figures and the suppression of events, irrelevant to bird activity.

In Fig. 7 we are interested in visualizing the benefit of using a U-net compared to the automatic segmentation in [6]. In 7 both the top row as well the bottom belong to spectrograms of sound scenes not containing bird activity but contaminated with noise. The whole spectrogram is rejected to have something related to bird activity. The bottom row in Fig 7 has even more noise and something sounding as operational artifact of the recorder. Again most of the audioscene is rejected as having potential bird activity.

Lastly, we examine the typical case that is of interest when scanning large volumes of audio where nothing interesting happens for large periods of time, and a case of interest is an outlier lost in the data. In Fig. 8 we see such a case of a vocalization in the presence of strong low-pass noise. Again the event is perfectly localized in the time-frequency spectrum, while all other sources are suppressed.

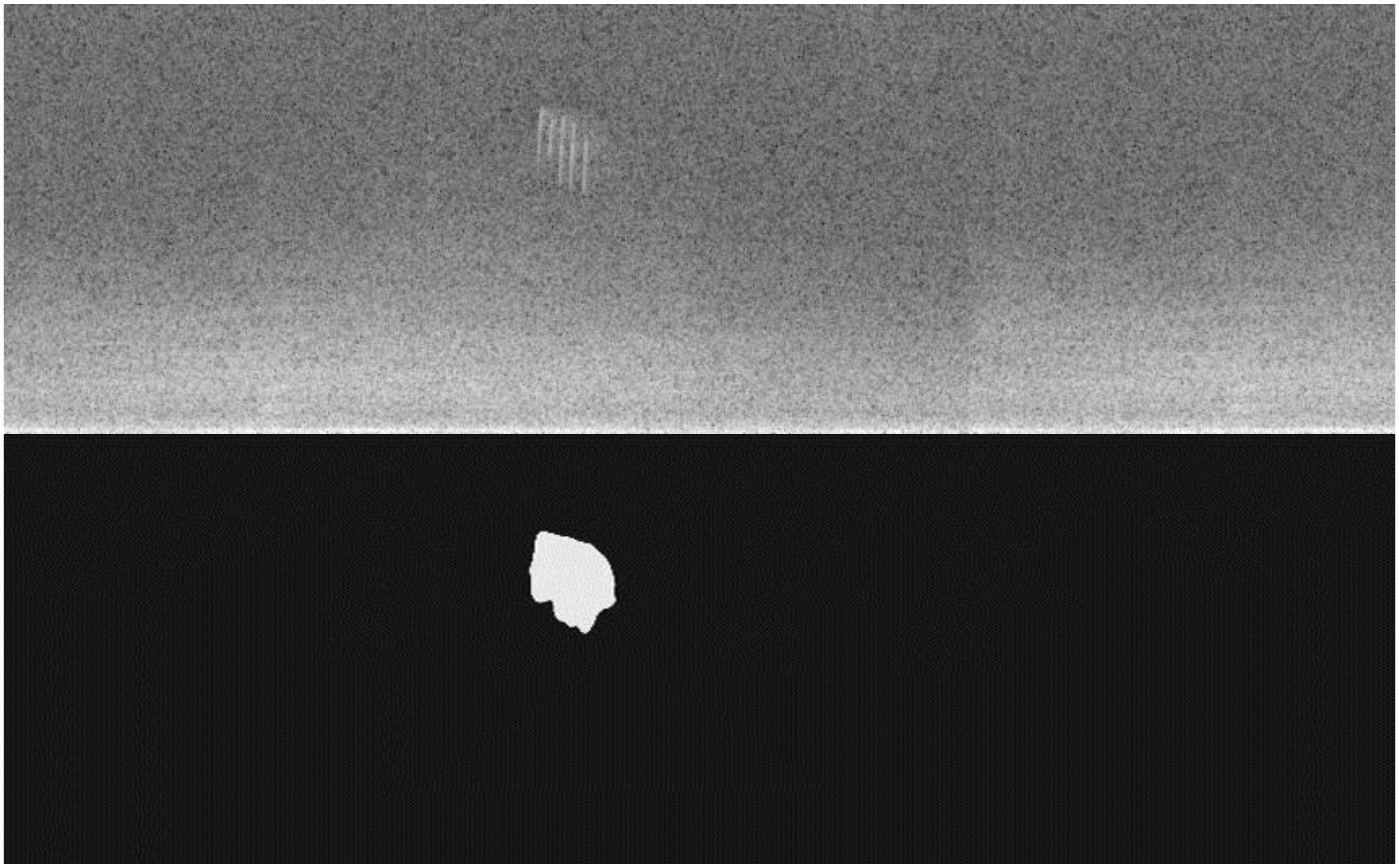

Fig. 8 An audio-scene with a sparse event of bird activity in the presence of strong wind and background traffic noise. (Up) Spectrogram, (Down) Binary mask.

## VI. Conclusion

In [23], the challenges of our time related to bird detection are thoroughly presented. This work as well, recognizes the necessity of processing big audio-data recorded in the field in order to extract the information needed to take decisions on biodiversity issues. We present a generic bird detection approach that crawls overnight over large number of recordings to examine the underlying bioacoustic activity. The core idea is that there is something common between images of X-rays, ultrasound electron microscopy and spectrograms. They are all 2D representations where the target appears as a blob that stands out from its background, and, therefore the very successful paradigm of Deep-learning with all its variations is a choice that calls to be investigated. It rests as future work to examine to what extent our approach is expandable to dealing with more refining questions such as: Can we fine-tune the U-net to discern a target species among all competing bird species in a soundscene? Can we suppress all kinds of acoustic interference just by including it in the training database? Definitely, more steps are needed in order to clarify these questions and to quantify the U-Net's efficacy in bird call/syllable segmentation.

## ACKNOWLEDGMENTS

We gratefully acknowledge the support of NVIDIA Corporation with the donation of a TITAN-X GPU partly used for this research. We used the Keras Deep learning library [24] in CUDA-CuDNN GPU mode. Python code in Anaconda Python 2.7.11 running in Ubuntu Linux environment and the Linux version of Matlab 2014b. We acknowledge the use of fragments of code of U-net found open on internet, and in several programming languages such as: http://lmb.informatik.uni-freiburg.de/people/ronneber/u-net/ among many others. The audioscenes void of bird vocalizations were recorded during the EC-funded project AMIBIO LIFE08-NAT-GR-000539.

## APPENDIX

Layers' composition in different versions of U-net: (Left): Simple U-net, (Middle): Additional Conv-pool layers added to the simple U-net and max-norm regularization in weight, (Right) Inception block inserted in layers instead of simple VGG-type conv-pool layers.

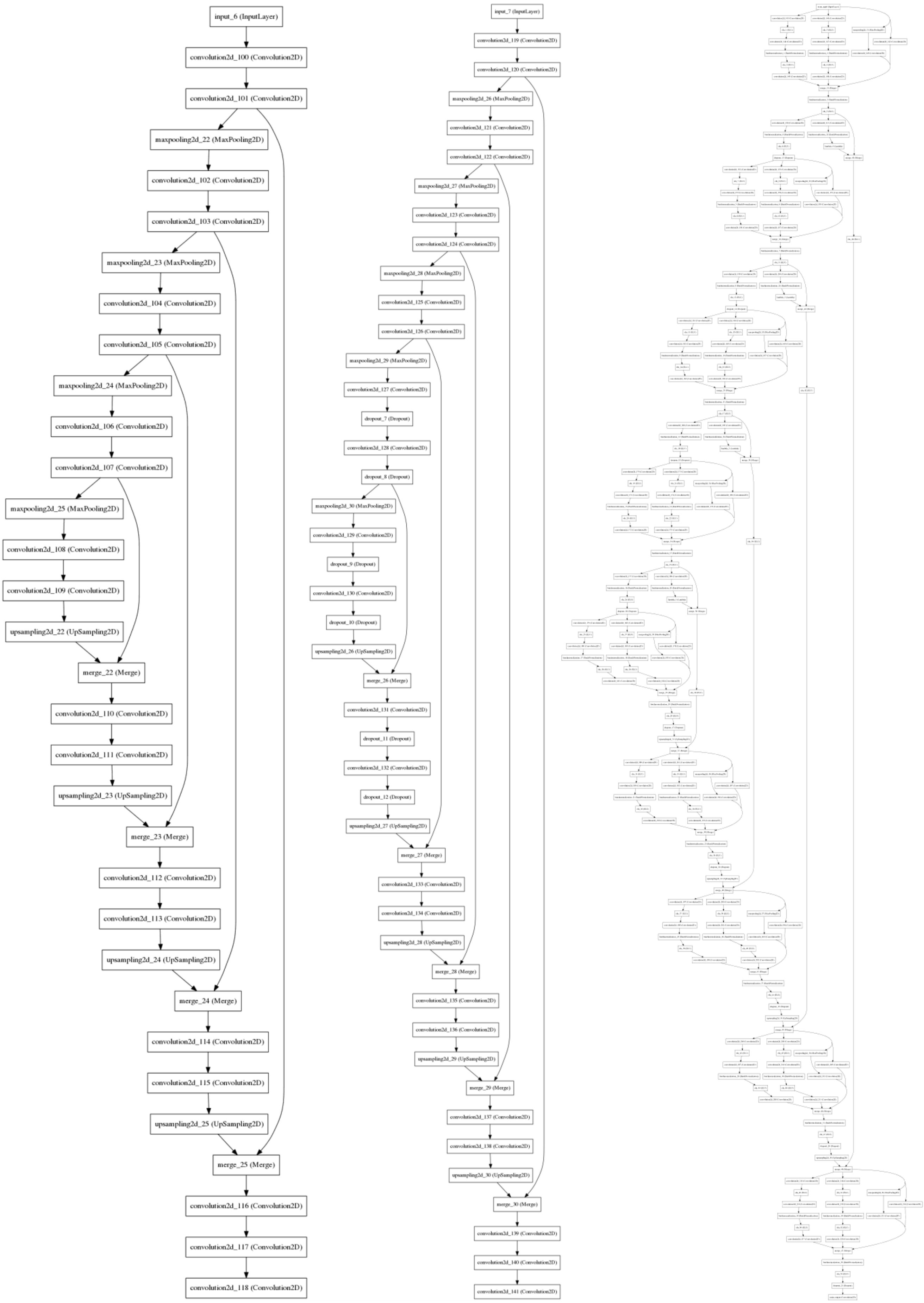

REFERENCES

[1] Catchpole C., and Slater P., "Bird Songs: Biological Themes and variations", Cambridge, 2008.

[2] P. Marler, "Bird calls: a cornucopia for communication," in Nature's Music: The Science of *Birdsong*, edited by P. Marler and H. Slabbekoorn, Chap. 5, pp. 132–177. New York, NY: Elsevier Academic Press, 2004.

[3] L. Baptista, and D. Kroodsma, 2001. Avian bioacoustics, Handbook of the Birds of the World, vol. 6: Mousebirds to Hornbills (J. del Hoyo, A. Ellio , and J. Sargatal, Eds.), Lynx Editions, Barcelona, Spain, pp. 11–52, 2001.

[4] J. Cai, D. Ee, B. Pham, P. Roe, and J. Zhang, "Sensor network for the monitoring of ecosystem: Bird species recognition," in 3rd International Conference on Intelligent Sensors, Sensor Networks and Information, (2008), pp. 293–298.

[5] Potamitis I., Ntalampiras S., Jahn O, Riede K., Automatic bird sound detection in long real-field recordings: Applications and tools, Applied Acoustics, Volume 80, June 2014, Pages 1-9, ISSN 0003-682X, http://dx.doi.org/10.1016/j.apacoust.2014.01.001.

[6] Lasseck M (2015). Towards Automatic Large-Scale Identification of Birds in Audio Recordings, Experimental IR Meets Multilinguality, Multimodality, and Interaction, Volume 9283 of the series Lecture Notes in Computer Science pp 364-375, Springer.

[7] P. Jancovic and M. Kokuer, "Automatic detection and recognition of tonal bird sounds in noisy environments," Journal of Advanced Signal Processing, 2011, 1–10, (2011)

[8] J. Kogan, and D. Margoliash, "Automated recognition of bird song elements from continuous recordings using dynamic time warping and hidden Markov models: a comparative study," Journal of the Acoustical Society of America, **103(4)**, 2185–2196, 1998.

[9] C. Kwan, K. Ho, G. Mei, Y. Li, Z. Ren, R. Xu, Y. Zhang, D. Lao, M. Stevenson, V. Stanford, and C. Rochet, "An automated acoustic system to monitor and classify birds," EURASIP Journal on Applied Signal Processing, Article ID 96706, 2006.

[10] S. Fagerlund, "Bird species recognition using support vector machines," EURASIP Journal on Applied Signal Processing, Article ID 38637, 2007.

[11] V. Trifa, A. Kirschel, C. E. Taylor, and E. E. Vallejo, "Automated species recognition of antbirds in a Mexican rainforest using hidden Markov models," Journal of the Acoustical Society of America, **123(4)**, 2424-2431, 2008.

[12] Y. Ren, M. Johnson, P. Clemins, M. Darre, S. Glaeser, T. Osiejuk, "A Framework for Bioacoustic Vocalization Analysis Using Hidden Markov Models, Algorithms," Algorithms **2(4)**, 1410-1428, 2009.

[13] F. Briggs, X. Fern, J. Irvine, "Multi-Label Classifier Chains for Bird Sound", Proceedings of the 30th International Conference on Machine Learning, Atlanta, Georgia, USA, 2013. JMLR, W&CP volume 28.

[14] F. Briggs, B. Lakshminarayanan, L. Neal, X. Fern, R, Raich, S., Hadley, and M. Betts, "Acoustic classification of multiple simultaneous bird species: A multi-instance multi-label approach," The Journal of the Acoustical Society of America, 131:4640, 2012.

[15] Potamitis I (2014) Automatic Classification of a Taxon-Rich Community Recorded in the Wild. PLoS ONE 9(5): e96936. doi:10.1371/journal.pone.0096936

[16] Potamitis I., Unsupervised dictionary extraction of bird vocalisations and new tools on assessing and visualising bird activity, Ecological Informatics, Volume 26, Part 3, March 2015, Pages 6-17, ISSN 1574-9541, http://dx.doi.org/10.1016/j.ecoinf.2015.01.002.

[17] Hendrik Vincent Koops, Jan van Balen, Frans Wiering, Automatic Segmentation and Deep Learning of Bird Sounds Experimental IR Meets Multilinguality, Multimodality, and Interaction, Volume 9283 of the series Lecture Notes in Computer Science pp 261-267, 2015

[18] Ronneberger, O., Fischer, P., Brox, T.: U-net: Convolutional networks for biomedical image segmentation. In: Medical Image Computing and Computer-Assisted Intervention–MICCAI 2015, pp. 234–241. Springer (2015)

[19] M. Towsey et al., "A toolbox for animal call recognition," Bioacoustics, vol. 21, no. 2, pp. 107–125, 2012.

[20] A. G. de Oliveira et al., "Bird acoustic activity detection based on morphological filtering of the spectrogram," Applied Acoustics, vol. 98, pp. 34–42, 2015.

[21] Szegedy Christian, Liu Wei, Jia Yangqing, Sermanet Pierre, Reed Scott, Anguelov Dragomir, Erhan, Dumitru, Vanhoucke, V, Rabinovich Andrew. Going Deeper with Convolutions, arXiv:1409.4842, 09-2014.

[22] Kaiming He, Xiangyu Zhang, Shaoqing Ren, Jian Sun. Deep Residual Learning for Image Recognition, arXiv:1512.03385v1, 2015

[23] Dan Stowell, Mike Wood, Yannis Stylianou, Hervé Glotin. Bird detection in audio: a survey and a challenge, arXiv:1608.03417 [cs.SD], 2016.

[24] https://github.com/fchollet/keras